\documentstyle[12pt]{article}
\begin{document}
\font\bss=cmr12 scaled\magstep 0
\title{Discrete symmetry's  chains and links between integrable
equations}
\author{A.V. Yurov\\
\small	Theoretical Physics Department,\\
\small	Kaliningrad State
University, \\
\small 236041, Al.Nevsky St., 14, Kaliningrad, Russia \\
\small	 email yurov@freemail.ru}
\date {}
\renewcommand{\abstractname}{\small Abstract}
\maketitle
\begin{abstract}
The discrete symmetry's dressing chains of the nonlinear Schr\"odinger
equation (NLS) and Davey-Stewartson equations (DS) are consider.
The modified NLS (mNLS) equation and the modified DS (mDS) equations
are obtained. The explicitly
reversible B\"acklund auto-transformations for the mNLS and mDS equations
are constructed. We demonstrate  discrete symmetry's conjugate chains of
the KP and DS models. The two-dimensional
generalization of the $P_4$ equation are obtained.
\end{abstract}
\thispagestyle{empty}
\medskip
\section{Introduction}
Darboux transformations (DT) are very useful method to construct exact solutions
of integrable PDE [1]. Some times ago a new approach to the DT (called dressing
chains of discrete symmetries) emerged at the horizon of soliton's mathematics.
This approach to proliferate integrable equations was proposed in the example of
the Korteweg-de Vries (KdV) and sine-Gordon equations [2]. The scheme
to proliferate the integrable equation (we choose the KdV for definitenes) starts
out from $LA$ pair and DT of this equation. After very simple transformations
(see Sec. 2) we obtain new integrable equations with  their $LA$  pairs. The
modified KdV (mKdV=$\rm m^1$KdV), $\rm m^2$KdV and $\rm m^3$KdV equations
were thus
constructed, the second of which can be reduced to the exponential
Calogero-Degasperis (CD) equation by an exponential point change of
variables and the third contains the elliptic CD equation. The only price we
have to pay for this simplicity is a rapidly growing amount of calculation.

Despite its apparent simplicity, the method of dressing chains is an extremely
powerful method, as can be illustrated in following example (see details in
[2-4]): as noted in [2] and [3], the $\rm m^n$KdV equations (n=0,...,3),
together
with the Krichever-Novikov equation, exhaust all the integrable equations of the
form $u_t+u_{xxx}+f(u_{xx},u_x,u)=0$. In [4] we generalized the dressing-chain
method to considerably (1+2)-dimensional nonlinear equations: the
Kadomtsev-Petviashvili (KP) and the Boiti-Leon-Pempinelli (BLP) equations. A new
result, which is characteristic precisely of multidimensional equations,
is the discovery of two types of dressing chains for the KP equation (we call
these the conjugate chains). The new chain can be constructed with the help of
binary DT.

The aim of this work is the generalization of the dressing chains method for
the nonlinear Schr\"odinger (NLS) and Davey-Stewartson (DS) equations. The plan of
this paper is as follows. In Sections 2 and 3 we review the dresing
chains method for
the higher and lower KdV equations and for the higher KP equations. The dressing
chains for the NLS and for the DS equations is discussed in Sections 4, 5. In
Sec. 6 we develop a technique for constructing of exact localized solution
of DS equations with a reduction constraint imposed and give some of these
solutions. We conclude with a discussion on
possible role of discrete symmetry's  dressing chains in the theory of integrable
PDE.

And last but not least. The dressing chains method allow us to
construct new integrable equations and, in the other hand, to
find a links between known ones.  For example, the sine-Gordon, KdV, and
CD equations can be constructed starting out from the KdV
equation via dressing chains.
{\em Maybe it is only one integrable equation really?}

\section{2D-$\rm P_4$ equation}

Let us begin our discussion of the Borisov-Zykov method by analysing
the simplest of all possible integrable systems, the KdV equation.
The starting point is the $LA$ pair for the KdV equation

$$
\psi_{xx}=(u-\lambda)\psi,\qquad
\psi_t=2(u+2\lambda)\psi_x-u_x\psi,
\eqno(1)
$$
where the KdV equation
$$
u_t-6uu_x+u_{xxx}=0.
\eqno(2)
$$
Setting $\tau=\phi_x/\phi$, where $\phi$ is a partial solution of (2) with
the $\lambda=\mu$ we can write the $LA$ pair as
$$
\tau_x=-\tau^2+u-\mu,\qquad
\tau_t=\left[2(u+2\mu)\tau-u_x\right]_x.
\eqno(3)
$$
Exepting $u$ from (4) we get
$$
\tau_t=6\tau^2\tau_x-\tau_{xxx}+6\mu\tau_x.
\eqno(4)
$$
If $\mu=0$ then we have the well-known mKdV equation. To construct
$LA$ pair for the (4) we  use the DT,
$$
u_1=u-2\tau_x,\qquad \psi_1=\psi_x-\tau\psi.
$$
Setting $\sigma=\psi_{1,x}/\psi_1$, we get the x-chain
$$
\left(\sigma+\tau\right)_x=-\sigma^2+\tau^2-\lambda+\mu.
\eqno(5)
$$
and t-chain,
$$
\left(\sigma+\tau\right)_t=
\left[2(-\tau_x+\tau^2+\mu+2\lambda)\sigma-2\tau\tau_x+6\mu\tau+2\tau^3\right]_x.
$$
At last, after introduction of the auxiliary function (denoted by $\Psi$),
$$
\sigma+\tau=\Psi,
$$
we obtain the $LA$ pair for the (4)
$$
\Psi_x=-\Psi^2+2\tau\Psi-\lambda+\mu,\qquad
\Psi_t=2\left[(\tau^2-\tau_x+\mu+2\lambda)\Psi+2(
\mu-\lambda)\tau\right]_x.
\eqno(6)
$$
Starting out from the (6) we can find the  $\rm m^2$KdV (which
be reduced to the exponential CD equation by an
exponential point change of variables) with it's $LA$ pair. Continuining
this procedure, we obtain the elliptic CD equation ($\rm m^3$KdV) and
{\em it is the end of this way}. The $\rm m^3$KdV equation is the last
equation which can be obtained by this method.

At first, we may wonder why the limitation appears by the n=3. However, we
can write the $LA$ pairs for the $\rm m^n$KdV with $n<4$ as two
Riccati equations. It is incorrectly for the  $\rm m^4$KdV [2]. This equation,
therefore, is different from the ones found earlier and cannot be used
by the old scheme. In the other hand, as we have mentioned in Introduction,
the $\rm m^n$KdV equations (n=0,...,3), together
with the Krichever-Novikov equation, exhaust all the integrable equations of the
form $u_t+u_{xxx}+f(u_{xx},u_x,u)=0$. So this limitation is something behind
that. We can understand the triviality of this limitation with the help of
the Painleve property. Let  consider the $\rm m^1$KdV equation (4).
The function $\tau$ is the solution of both this equation and two
Riccati equations (3) (inserting $\tau_x$ from the  first equation (3) in
the second one we obtain the second Riccati equation), so it is obvious that
the function $\tau$ has the Painleve property and that it is incorrectly for
the solutions of the $\rm m^3$KdV with $n>3$.

In [5] the continuation of the KdV equation hierarchy in direction corresponding
to the negative power of the spectral parameter is constructed by the following
way: let the L-operator has the form (1),
$$
L=-\partial_x^2+u(x,t),
$$
and the nonlinear equations  are
$$
L_t=[L,A_{_N}].
$$
The operators $A_{_N}$ have the form
$$
A_{_N}=\sum_{m=-1}^{N} K_m(L)^m,
$$
where $N=-1,\,\,-2,\,\,...$ and $K_m$ are some operators.

The first lower KdV (N=-1) equation has the form (see [5]),
$$
{\rm KdV}_{-1}(\sigma)\equiv (\sigma_x^2+\sigma_{xx})_t-\left(e^{2\sigma}\right)_x=0,
$$
where $\sigma$ is connected with $u$ by the following way
$$
u=-\sigma_{xx}-\sigma_x^2.
$$
Let $\sigma=iq/2$, where $q=q(x,t)$. It easy to see that
$$
{\rm KdV}_{-1}(\sigma)=\frac{1}{2}\left(i\partial_x-q_x\right)(q_{xt}-2\sin  q)=0,
\eqno(7)
$$
so we have the Miura transformation between  the $\rm KdV_{-1}$ equation and
the
sine-Gordon equation.

LA-pair for the (7) has the form,
$$
\psi_{xx}=\left(\frac{iq_{xx}}{2}-\frac{q_x^2}{4}+\lambda^2\right)\psi,\qquad
\psi_t=\frac{1}{2\lambda^2}e^{iq}
\left(\psi_x-\frac{iq_x}{2}\psi\right),
\eqno(8)
$$
Starting out from the (8)  we can  find the LA-pair for the
sine-Gordone
equation. To do this we introduce new function $\tilde\psi$
$$
\tilde\psi=\frac{1}{\lambda}\left(\psi_x-\frac{i}{2}q_x\psi\right).
\eqno(9)
$$
therefore
$$
\tilde\psi_x=\lambda\psi-\frac{i}{2}q_x\tilde\psi,\qquad
\psi_t=\frac{1}{2\lambda}e^{iq}\tilde\psi,\qquad
\tilde\psi_t=-\frac{1}{2\lambda}\left(iq_{xt}-e^{iq}\right)\psi.
\eqno(10)
$$
Substituting $2\sin q$ in place of $q_{xt}$ (in the eq. for the
$\tilde\psi_t$) we can see that  (9), (10) are the well known LA-pair for the
sine-Gordon equation.

Thus, we can use the method of dressing chains to both higher and
lower KdV equations. The sine-Gordon, KdV, and CD equations can be constructed
starting out from the KdV equation via dressing chains of  discrete
symmetries. There are simply different representations of the same
equation. From this this standpoint, {\em the difference between sine-Gorgon,
KdV, mKdV and CD equations is similar to the difference between the
Maxwell equations in different gauges!}

Surprisingly, much of the analysis of the simple system carries over
directly to the multidimensional case.
The KP equations are given by
$$
u_t+6uu_x+u_{xxx}=-3\alpha^2\,v_y,\qquad
v_x=u_y,
$$
where $u=u(x,y,t)$, $v=v(x,y,t)$, $\alpha^2=\pm 1$. The $LA$ pair for the KP
equations is:
$$
\alpha\psi_y+\psi_{xx}+u\psi=0,\qquad
\psi_t+4\psi_{xxx}+6u\psi_x+3(u_x-\alpha v)\psi=0.
\eqno(11)
$$
Setting $f\equiv \log\phi$,\,$\tau\equiv f_x$,
where $\phi$ is a partial solution of (11), we can write $LA$ pair (11) as
$$
\begin{array}{l}
\alpha\tau_y+\left(\tau_x+\tau^2+u\right)_x=0\\
\\
\tau_t+\left(4\tau_{xx}+12\tau\tau_x+4\tau^3+6u\tau+3u_x-3\alpha v\right)_x=0.
\end{array}
\eqno(12)
$$
It follows from the first equation in (12) that
$$
u=-\tau_x-\tau^2-\alpha F,\qquad
F_x=\tau_y.
\eqno(13)
$$
Substituting (12) into the second equation in (13) and making some
simple transformations, we obtain the well-known mKP equation [6]
$$
\tau_t-6\tau^2\tau_x+\tau_{xxx}=3\alpha(2\tau_x F-\alpha F_y),\qquad
F_x-\tau_y=0.
\eqno(14)
$$
We now construct dressing chains of discrete symmetries. The KP equation admits
the Darboux transformation [1]
$$
\psi_1=\psi-\tau\psi,\qquad
u_1=u+2\tau_x,\qquad
v_1=v+2\tau_y.
\eqno(15)
$$
Setting $s\equiv\log\psi_1$, $\sigma=s_x$, we see that $\sigma$ satisfies
the system of equations obtained from (12) by replacing
$u\to u_1$ and $v\to v_1$. Comparing these equations with (12) and eliminating
the potentials $u$ and $v$, we obtain
$$
\begin{array}{l}
\alpha(s-f)_y+(s+f)_{xx}+s_x^2-f_x^2=0,\\
\\
(s-f)_t+\left[2(2s+f)_{xx}+6s_x^2-3f_x^2\right]_x+4s_x^3+
6\left(f_{xx}-f_x^2-\alpha f_y\right)s_x-\\
\\
6\alpha(f_{xy}-f_xf_y)+2f_x^3=0.
\end{array}
\eqno(16)
$$
The first equation in (16) determines the $y$- chain and the second
determines the $t$-chain, denoted by $C_y^{(+)}$ and $C_t^{(+)}$
respectively. Later (in Sec. 3) we will generalize the Borisov-Zykov
approach for the case of the KP equation, and we will find that the results
of Borisov-Zykov carry over with only one important changes.

As we seen, the Painleve property is connected with the dressing chains.
The aproach of using dressing chain (5) to construct the $P_4$ equation
was proposed in [7]. It is interesting to apply this method to the chain
(16).

To do this it is helpful to incert the periodic condition.
Well known that the periodic condion for the (5) led to the
$P_4$ equations. Let consider the dressing chain $C_y^{(+)}$ (16).
Setting $f=f_n$, $s=f_{n+1}$ we use the condition of periodic
$$
f_{n+N}=f_n+c(y).
\eqno(17)
$$
It easy to see that
$$
u_{n+N}=u_n+\alpha c',
$$
so the condition (17) get us some generalization of the harmonic oscillator.
We choose $N=3$ and $c=-2y/\alpha$. Introducing new fields $g_n$,
$n=1,2,3$
$$
f_1=\frac{1}{2}\left(g_1-g_2+g_3+c\right),\qquad
f_2=\frac{1}{2}\left(g_1+g_2-g_3-c\right),\qquad
f_3=\frac{1}{2}\left(-g_1+g_2+g_3+c\right),
$$
we can now rewrite the equations for the $f_n$ by the following way
\[
{ \alpha}\, \left( \! \,{{\it g_2}_{{y}}} - {{\it g_3}_{{y}}} - {{c
}_{{y}}}\, \!  \right)	+  \left( \! \,{{\it g_2}_{{x}}} - {{\it
g_3}_{{x}}}\, \!  \right) \,{{\it g_1}_{{x}}} + {{\it g_1}_{{\it xx}
}}=0,
\]
\[
{ \alpha}\, \left( \! \, - {{\it g_1}_{{y}}} + {{\it g_3}_{{y}}} +
{{c}_{{y}}}\, \!  \right)  +  \left( \! \, {{\it g_3}_{{x}}} - {
{\it g_1}_{{x}}}\, \!  \right) \,{{\it g_2}_{{x}}} + {{\it g_2}_{
{\it xx}}}=0,
\]
\[
{ \alpha}\, \left( \! \,{{\it g_1}_{{y}}} - {{\it g_2}_{{y}}} - {{c
}_{{y}}}\, \!  \right)	+  \left( \! \,{{\it g_1}_{{x}}} - {{\it
g_2}_{{x}}}\, \!  \right) \,{{\it g_3}_{{x}}} + {{\it g_3}_{{\it xx}
}}=0.
\]
Excluding $g_3$, and using the	compatibility condition
$D_y g_{2,xx}=D_x^2 g_{2,y}$ we get  the nonlinear
equation which can be written via some transformation as
$$
\begin{array}{l}
\displaystyle{
z_{xx}=\frac{1}{2}\frac{z_x^2}{z}+\frac{3}{2}z^3+4xz^2+2(x^2-2)z+
\frac{\beta}{z}+}\\
\\
\displaystyle{
+\frac{3\alpha^2q^2}{{2z}}-3\alpha qz+
\frac{\alpha}{z}D_x^{-1}D_y\left(z^3+2xz^2-3\alpha qz\right),\qquad
z_y=q_x.}
\end{array}
\eqno(18)
$$
where $z={{\it g_{1}}_{{x}}}$ and $q={{\it g_{1}}_{{y}}}$.
The one-dimensional limit (where $D_y=q=0$) of Eq. (18) is the $P_4$
equation, so
the (18) is the two-dimensional generalization of the $P_4$ equation and
it should be called the 2D-$\rm P_4$ equations.

The price we pay for the D=2 is that locality in $x$ is lost. That is in
the usual run of things for the multidimensional integrable systems.

\section{Conjugate chains for the KP equations}

The correspondence between the KdV dressing chains formalism that we
develop in the previous section and the KP dressing chains formalism
is quite remarkable. We find that almost the entire KdV formalism
can be imported into the KP chains.

Chains (16) involve two $LA$ pairs for Eqs. (14), which arise after
introducting the auxiliary function (denoted by $\Psi$) according to one of
two rules,
$$
\Psi=s-f,\qquad \Psi=s+f.
\eqno(19)
$$
Using the first rule, we obtain
$$
\begin{array}{l}
\alpha\Psi_y+\left(2f+\Psi\right)_{xx}+2f_x\Psi_x+\Psi_x^2=0,\\
\\
\Psi_t+2\left(2\Psi_{xx}+6f_x\Psi_x+3\Psi_x^2+3\theta\right)_x+
6\theta\Psi_x+4\left(3f+\Psi\right)_x\Psi_x^2=0,\\
\\
\theta\equiv f_{xx}+f_x^2-\alpha f_y
\end{array}
\eqno(20)
$$
from (16). The compatibility condition for these equations is
$$
\left(f_t-2f_x^3+f_{xxx}\right)_x=3\alpha\left(2f_{xx}f_y-\alpha f_{yy}\right),
\eqno(21)
$$
which is merely another form of the mKP equation (for $\tau=f_x$ and
$F=f_y$).

This process can be repeated. In particular, the system (20) contains a new
nonlinear equation obtained by eliminating the potential $f$. For this purpose,
we linearize the first equation in (20) by the substituting
$$
f=-\frac{1}{2}\left(\Psi+\alpha\xi\right),
\eqno(22)
$$
where $\xi=\xi(x,y,t)$. Inserting (22) in the second equation in (20),
differentiating it with respect to $x$, and introducing $S\equiv \Psi_x$,
we obtain the system
$$
\begin{array}{l}
S_t+S_{xxx}-\frac{3}{2}S^2S_x=-3\alpha^2
\left[\left(\frac{1}{2}\xi_x^2+\xi_y\right)S
+\left(\frac{1}{2}\xi_x^2+\xi_y\right)_x\right]_x,\\
\\
S_y=\left(\xi_{xx}+\xi_xS\right)_x,
\end{array}
\eqno(23)
$$
Although Eqs. (23) look like a two-dimensional generalization of the
mKdV equation, the one-dimensional limit
(where $\partial_y=0$) of (20) is
$$
g_t+\left(g_{xx}-\frac{\alpha^2}{2}g^3-\frac{3}{2}\frac{g_x^2}{g}
\right)_x=0,
$$
(where $g(x,t)\equiv\xi_x=\exp(-\Psi)$) rather than the mKdV equation. After
an exponential point change of variable, this equation reduced to the
exponential CD equation. Therefore, Eq. (23) should be called the
two-dimensional CD equation rather than the two-dimensional mKdV equation.

In [2], another representation for $\Psi$ was used in the derivation of the
CD equation, namely, the representation determined by the second formula
in (19). We have used this representation in [4].

In addition to the usual DT, the KP equations
admit the so-called binary Darboux transformations [1], which, as we
now show, allow us to construct new KP dressing chains.

It is obvious that the KP equation admits the $LA$ pair
$$
-\alpha\chi_y+\chi_{xx}+u\chi=0,\qquad
\chi_t+4\chi_{xxx}+6u\chi_x+3(u_x+\alpha v)\chi=0,
\eqno(24)
$$
which can be obtained from (11) by simple replacing
$\alpha\to - \alpha$. Let
$$
\begin{array}{l}
\displaystyle
{dQ(\chi,\psi)=\chi\psi dx+\frac{1}{\alpha}\left(\chi_x\psi-\chi\psi_x\right)dy+
4\left(\psi_x\chi_x-\chi_{xx}\psi-\chi\psi_{xx}-\frac{3}{2}u\chi\psi\right)dt}\\
\\
Q(\chi,\psi)\equiv\int dQ(\chi,\psi).
\end{array}
$$
It is easy to see that the one-form $dQ(\chi,\psi)$ is closed if
$\psi$ and $\chi$ are solutions of Eqs. (11) and (24). Pair (24)
also admits the Darboux transformation
$$
\chi_{-1}=\chi-\rho\chi,\qquad
u_{-1}=u+2\rho_x,\qquad
v_{-1}=v+2\rho_y,
\eqno(25)
$$
where $\rho=(\log{\tilde\chi})_x$, and ${\tilde\chi}$ is a partial
solution of (24). Applying DT (15), we have
$$
\chi_1=\frac{A+BQ(\chi,\psi)}{\psi},
$$
where $A$ and $B$ are arbitrary constants. Now using (25), we obtain
$$
u_{1,-1}=u+2\left[\log(A+BQ(\chi,\psi))\right]_{xx},\qquad
v_{1,-1}=v+2\left[\log(A+BQ(\chi,\psi))\right]_{xy}.
$$
We set $\sigma\equiv \chi_{1,x}/\chi_1$. This function satisfies the
equation obtained from (12) by replacing
$u\to u_1$, $v\to v_1$ and $\alpha\to-\alpha$. In other words, this yield
new integrable $y$- and $t$-chains ($C_y^{(-)}$ and
$C_t^{(-)}$), which can be called {\em conjugate} to the $C^{(+)}$ chains
discussed above:
$$
\begin{array}{l}
-\alpha(s+f)_y+(s+f)_{xx}+s_x^2-f_x^2=0,\\
\\
(s+f)_t+\left[4(s+f)_{xx}+6s_x^2-3f_x^2\right]_x+4s_x^3+
6\left(f_{xx}-f_x^2-\alpha f_y\right)s_x-6\alpha f_xf_y-2f_x^3=0,
\end{array}
$$
where we still have $\tau=f_x$ and $\sigma=s_x$. Comparing the expressions
for $C^{(-)}$ and $C^{(+)}$ from (16), we see that these chains do not reduce
to each other, thereby justifying the term "conjugate". This implies that
the chains $C^{(-)}$ lead to new integrable equations and also to the
corresponding $LA$ pairs. To construct these, we should again use two
possibilities to define $\Psi$, see Eq. (19). In the present case, it is
convenient to choose the second formula. As the result, we obtain yet another
$LA$ pair for the mKP equation (21):
$$
\begin{array}{l}
-\alpha\Psi_y+\Psi_{xx}-2f_x\Psi_x+\Psi_x^2=0,\\
\\
\Psi_t+2\left(2\Psi_{xx}-3f_x\Psi_x+3\Psi_x^2\right)_x+
6\left(f_x^2-\alpha f_y\right)\Psi_x-6f_x\Psi_{xx}+
4\left(\Psi-3f\right)_x\Psi_x^2=0.
\end{array}
$$
Linearizing the first equation by the substitution that
differs from the (22) only by the sign of the right-hand
side and defined $S\equiv\Psi_x$, we get
$$
\begin{array}{l}
S_t+4S_{xxx}-\frac{3}{2}S^2S_x=3\alpha
\left[\xi_{xx}S+2\xi_xS_x+
\alpha\left(\xi_y-\frac{1}{2}\xi_x^2\right)S\right]_x,\\
\\
\alpha S_y=\left(S_x-\alpha\xi_xS\right)_x.
\end{array}
\eqno(26)
$$

Equations (26) are similar to (23) and also reduce to the CD equation in
the one-dimensional limit (for $\alpha^2=1$). At the same time,
as follows from the method of their construction, Eqs. (26) do not reduce
to system (23) via a 1:1 change of dependent variables (by a
gauge transformation); therefore, Eqs. (26) can be called conjugate
to the Eqs. (23).

\section{Dressing chains for the NLS equation}

The results in last section indicates that generalizing the
dressing-chain method to considerably more complicated NLS and
DS equations can be very fruitful. In this section we use the
dressing chains of discrete symmetries to proliferate the
NLS equation. Recently, Shabat has studied such chains for the
Zakharov-Shabat spectral problem in some unusual gauge [8]. We have
another aim: to construct  higher mNLS equations. We can do it in
two gauges, each with its own advantages and disadvantages:
\newline
\newline
(1) {\em Standart symmetric Zakharov-Shabat gauge}. In this gauge we have
very clear connections between NLS and $\rm m^n$NLS equations. The disadvantage
of this gauge is a very rapidly growing ammount of calculations. For
the $n>2$ we obtain very cumbersome equations.
\newline
(2) {\em New Shabat gauge} [8]. The advantage of this gauge is that equations
has more compact form. However, the
connection between real NLS and real mNLS equations is not obvious in this
gauge. The reconstruction of the $\rm m^n$NLS for
a $n>2$ must be checked tediously.
\newline
\newline
Here, we choose the standard gauge for the spectral problem and therefore
the $LA$ pair for the NLS equation is
$$
\Psi_t=-2i\sigma_3\Psi\Lambda^2+2iU\Psi\Lambda+V\Psi,\qquad
\Psi_x=-i\sigma_3\Psi\Lambda+iU\Psi,
\eqno(27)
$$
where
$$
U=\left(\begin{array}{cc}
0&u\\
v&0
\end{array}\right),
\qquad
\sigma_3=\left(\begin{array}{cc}
1&0\\
0&-1
\end{array}\right),
\qquad
V=\sigma_3\left(iU^2-U_x\right),
$$
and $\Lambda$ is arbitrary constant matrix.
The  compatibility condition get us the NLS equations
$$
iu_t+u_{xx}+2u^2v=0,\qquad
-iv_t+v_{xx}+2v^2u=0.
$$
Let $\Phi$ is a partial solution of (27) with the
$\Lambda={\rm diag}((\lambda+\mu)/2,(\lambda-\mu)/2)$,
$\Psi$ is the same for the
$\Lambda_1={\rm diag}((\lambda_1+\mu_1)/2,(\lambda_1-\mu_1)/2)$ and
$\tau\equiv \Phi\Lambda\Phi^{-1}$. This matrix function is the solution of the
system
$$
\tau_x=i[\tau,\sigma_3]\tau+i[U,\tau],\qquad
\tau_t=2\tau_x\tau+[V,\tau].
\eqno(28)
$$
The Darboux transformation has the form
$$
\Psi\to\Psi_1=\Psi\Lambda_1-\tau\Psi,\qquad
U\to U_1=U+[\tau,\sigma_3].
\eqno(29)
$$
There are two discrete symmetries of the Zakharov-Shabat spectral problem:
S- and T-symmetries. The transformation (29) is $S^2$ symmetry. We must
use  squared S-symmetry because an elementary S-symmetries are
unsufficient to construct exact solutions of the NLS.

Using (29) we obtain
$$
u\to u_1=u-2b,\qquad v\to v_1=v+2c,
\eqno(30)
$$
where
$$
\tau=\left(\begin{array}{cc}
a&b\\
c&d
\end{array}\right).
\eqno(31)
$$
Substituting (31) into (28) we get
$$
\begin{array}{c}
a_x=-d_x=-i(2bc-uc+vb),\qquad
b_x=-i(2bd+(a-d)u),\\
c_x=i(2ac+(a-d)v)
\end{array}
\eqno(32)
$$
and
$$
\begin{array}{c}
a_t=(a^2)_x+2b_xc-bv_x-cu_x,\qquad d_t=(d^2)_x+2bc_x+bv_x+cu_x\\
{}\\
b_t=2(ibuv+a_xb+b_xd)+(a-d)u_x,\qquad
c_t=2(-icuv+ac_x+cd_x)+(a-d)v_x.\\
\end{array}
\eqno(33)
$$
Calculating the determinant and trace of matrix $\tau$ we get
$$
ad-bc=\frac{\lambda^2-\mu^2}{4},\qquad a+d=\lambda.
\eqno(34)
$$
Using (34) we get
$$
a=\frac{1}{2}\left(\lambda\pm\sqrt{\mu^2-4bc}\right).
\eqno(35)
$$
Eliminating $u$ and $v$ from the last two Eqs. (32) and putting everything
together we find mNLS ($\rm m^1$NLS) equation
$$
\begin{array}{c}
\left(\mu^2-4bc\right)\left(ib_t+b_{xx}-2b^2c\right)+
2\lambda\left(\lambda c+2ic_x\right)b^2+
2(b_xc+2bc_x)b_x\equiv \left(\mu^2-4bc\right)\beta(b,c)=0,\\
{}\\
\left(\mu^2-4bc\right)\left(-ic_t+c_{xx}-2c^2b\right)+
2\lambda\left(\lambda b-2ib_x\right)c^2+
2(bc_x+2b_xc)c_x\equiv	\left(\mu^2-4bc\right)\gamma(b,c)=0.
\end{array}
\eqno(36)
$$
The designeshions $\beta(b,c)$ and $\gamma(b,c)$ will be useful
(see (41), (42)).

As we have mentioned above, the NLS equation has a broader set of
symmetries. In addition to the Darboux transformations, the discrete
symmetries include the Schlesinger transformations or T-symmetries.
The T-symmetries produce an explicity invertible B\"acklund
auto-transformations for the NLS equation
$$
u\to u_1=u_{xx}+u^2v-\frac{u_x^2}{u},\qquad v\to v_1=\frac{1}{u},
$$
and
$$
u\to u_{-1}=\frac{1}{v},\qquad
v\to v_{-1}=v_{xx}+v^2u-\frac{v_x^2}{v},
$$
where
$$
(u_1)_{-1}=(u_{-1})_1=u,\qquad
(v_1)_{-1}=(v_{-1})_1=v.
$$
These symmetries are related
to the Toda chain and some of its generalizations [9].

The mNLS equation (36) has the similar property.
We have very  cumbersome Schlesinger transformations for the (36) so
we show ones when $\lambda=\mu=0$. In this case, the (36) has an elegant form
$$
\begin{array}{l}
\displaystyle
{ib_t+b_{xx}-2b^2c-\frac{b_xc_x}{c}-\frac{1}{2}\frac{b_x^2}{b}=0,}\\
{}\\
\displaystyle
{-ic_t+c_{xx}-2bc^2-\frac{b_xc_x}{b}-\frac{1}{2}\frac{c_x^2}{c}=0.}
\end{array}
\eqno(37)
$$
The explicity invertible B\"acklund
auto-transformations for the (37) are
$$
\begin{array}{l}
\displaystyle
{b\to b_1=\frac{1}{4}
\frac{(2cb_x^2-2cbb_{xx}+c_xb_xb+4c^2b^3)^2}
{c^2b(b_x^2-4cb^3)},\qquad
c\to c_1=\frac{4cb^2}{b_x^2-4cb^3},}\\
{}\\
\displaystyle
{b\to b_{-1}=\frac{4bc^2}{c_x^2-4bc^3},\qquad
c\to c_{-1}=\frac{1}{4}
\frac{(2bc_x^2-2bcc_{xx}+b_xc_xc+4b^2c^3)^2}
{b^2c(c_x^2-4bc^3)}}.
\end{array}
$$
It is easy to verify that
$$
(b_1)_{-1}=(b_{-1})_1=b,\qquad
(c_1)_{-1}=(c_{-1})_1=c.
$$

To obtain dressing chains of discrete symmetries $S^2$ we construct
new matrix function $\tau_1=\Psi_1\Lambda_1\Psi_1^{-1}$ with $\Psi_1$ from
the (29). Its elements $a_1$, $b_1$, $c_1$ and $d_1$ are solutions
of  (32)-(35) by replacing
$\lambda\to\lambda_1$, $\mu\to\mu_1$, $u\to u_1$ and $v\to v_1$ (see (30)).
Eliminating potentials $u$ and $v$, we obtain our chains
$$
\begin{array}{l}
\displaystyle
{(\lambda-2a)b_{1,x}-(\lambda_1-2a_1)b_x+2i(\lambda_1-a_1)(\lambda-2a)b_1-
2i(\lambda_1-2a_1)ab=0},\\
\displaystyle
{(\lambda-2a)c_{1,x}-(\lambda_1-2a_1)c_x-2i(\lambda-2a)a_1c_1+2i(\lambda_1-2a_1)
(\lambda-a)c=0},\\
{}\\
\displaystyle
{b_{1,t}b-b_1b_t+2\left[(\lambda-a)b_1b_x-(\lambda_1-a_1)b_{1,x}b\right]+
(\lambda_1-2a_1)\left(b^2\right)_x+K_1b_1+M_1b=0},\\
\displaystyle
{c_{1,t}c-c_1c_t+2\left(ac_1c_x-a_1c_{1,x}c\right)+(\lambda_1-2a_1)\left(c^2\right)_x+
K_2c_1+M_2c=0},
\end{array}
\eqno(38)
$$
where
$$
\begin{array}{l}
K_1=4iC(2,2)b^2+2[(a-a_1)_x+2iB(-2,-2)c]b-(\lambda-2a)B_x(-2,-2),\\
{}\\
K_2=-4iB(0,2)c^2-2[(a-a_1)_x+2iC(0,-2)b]c-(\lambda-2a)C_x(0,-2),\\
{}\\
M_1=(\lambda_1-2a_1)B_x(-2,-2),\qquad M_2=(\lambda_1-2a_1)C_x(0,-2),\\
{}\\
\displaystyle
{B(n,k)=\frac{ib_x+(n\lambda-ka)b}{\lambda-2a},\qquad
C(n,k)=\frac{ic_x+(n\lambda-ka)c}{\lambda-2a}},
\end{array}
$$
and fields $a$, $d$ ($a_1$, $d_1$) are expressed in term of
$b$, $c$ ($b_1$, $c_1$) via (34)-(35).

Chains (38) involve $LA$ pair for the mNLS (36), which arise after introducing
the auxiliary fields
$$
\psi\equiv\frac{b_1}{b},\qquad \phi\equiv\frac{c_1}{c},\qquad A\equiv a_1.
$$
After some calculations we obtain
$$
\begin{array}{l}
\displaystyle{
\psi_x=\left(2i(A-\lambda_1)-\frac{b_x}{b}\right)\psi+\frac{i(2A-\lambda_1)}{b}
B(0,2),}\\
\displaystyle{\phi_x=\left(2iA-\frac{c_x}{c}\right)\phi+\frac{i(2A-\lambda_1)}{c}
C(2,2),}\\
{}\\
\displaystyle{\psi_t=2(\lambda_1-A)\psi_x-4ibc\phi\psi^2-2i(bC(2,2)\psi+cB(0,2)\phi)\psi+P\psi+
\frac{\lambda_1-2A}{b}B_x(0,2)},\\
\displaystyle{\phi_t=2A\phi_x+4ibc\psi\phi^2+2i(bC(2,2)\psi+cB(0,2)\phi)\phi+Q\phi-
\frac{\lambda_1-2A}{c}C_x(2,2)},\\
{}\\
A_x=-i[2bc\phi\psi+bC(2,2)\psi+cB(0,2)\phi],\\
A_t=-4i\lambda_1 bc\phi\psi-b[2iAC(2,2)+C_x(2,2)]\psi+
c[2i(A-\lambda_1)B(0,2)+B_x(0,2)]\phi,
\end{array}
\eqno(39)
$$
where
$$
\begin{array}{l}
\displaystyle
{P=\frac{[2b_x(\lambda_1-A)-ib_{xx}](\mu^2-4bc)+2\left(bc\right)_x[2b(\lambda-a)-ib_x]}
{b\left(\mu^2-4bc\right)},}\\
{}\\
\displaystyle
{Q=\frac{(2c_xA+ic_{xx})(\mu^2-4bc)+2\left(bc\right)_x(2ca+ic_x)}
{c\left(\mu^2-4bc\right)}}.
\end{array}
$$
The mNLS equation (36) arise from the compatibility condition of (39):
$$
\left(\psi_x\right)_t=\left(\psi_t\right)_x,\qquad
\left(\phi_x\right)_t=\left(\phi_t\right)_x,\qquad
\left(A_x\right)_t=\left(A_t\right)_x.
\eqno(40)
$$
From the first two equations (40) we get two nonlinear equations
$$
\beta_1(b,c)=0,\qquad \gamma_1(b,c)=0.
$$
The connection between these equations and the mNLS (36) is given by
$$
\beta_1(b,c)=b\left(\frac{\beta(b,c)}{b}\right)_x,\qquad
\gamma_1(b,c)=c\left(\frac{\gamma(b,c)}{c}\right)_x,
\eqno(41)
$$
so we have not (36) but
$$
\beta(b,c)=mb,\qquad \gamma(b,c)=m'c,
\eqno(42)
$$
where $m$ ¨ $m'$ are arbitrary constants. Substituting $b_t$ and $c_t$
(which are expressed from (42)) into the third equation of the system (40)
we obtain $m'=m$. Thus, using the gauge transformation
$$
b\to e^{-imt}b,\qquad
c\to e^{imt}c,
$$
we can reduce the (42) to mNLS (36)
$$
\beta(b,c)-mb\to e^{-imt}\beta(b,c)=0,\qquad
\gamma(b,c)-mc\to e^{-imt}\gamma(b,c)=0.
$$
This completes the proof that the system (39) is the $LA$ pair
of mNLS. We are giving
two formulas (for convenience of reader) which are useful to check of
this $LA$ pair:
$$
\displaystyle
{a_x=\frac{\left(bc\right)_x}{\lambda-2a},\qquad
a_t=\lambda a_x+b_xc-bc_x-bC_x(0,-2)+cB_x(-2,-2),}
$$

We will shortly see the advantage of carefully working out the details
of the NLS dressing chains. We will find that almost all of this
formalism carried over directly into the more complicated DS dressing
chains. Formulaes for the DS dressing chains are very bulki but ones
has the same structure.

\section{Conjugate chains of the DS equations}

The DS equations has the form
$$
\begin{array}{l}
\displaystyle
{iu_t+u_{xx}+\frac{1}{\alpha^2}u_{yy}-\frac{2}{\alpha^2}u^2v+qu=0,\qquad
-iv_t+v_{xx}+\frac{1}{\alpha^2}v_{yy}-\frac{2}{\alpha^2}v^2u+qv=0},\\
{}\\
q_{yy}-\alpha^2q_{xx}=-4\left(uv\right)_{xx},
\end{array}
\eqno(43)
$$
where $u=u(x,y,t)$, $v=v(x,y,t)$, $q=q(x,y,t)$. We have the DS-I system
if $\alpha=1$ and  DS-II, if $\alpha=i$. Under the reduction
$v=\pm \overline u$,
one obtains a known model that describes the propogation of a small amplitude
wave packet that is quasi-one-dimensional and quasi-monochromatic over the
surface of a nonviscous curl-free liquid [10]. The quantity $u$ is the envelope of
the velocity potential, while $q$ describes the nonlocal flow generated by
the wave packet. A different application of (43) is related to the
dynamics of plasma waves [11]. We woun`t use this reduction restriction
in this section.

$LA$ "pair" for the (43) has the form
$$
\begin{array}{l}
\psi_y=\alpha\psi_x+u\phi,\qquad
\phi_y=-\alpha\phi_x+v\psi,\\
{}\\
\displaystyle
{\psi_t=2i\psi_{xx}+\frac{2i}{\alpha}u\phi_x+
\left(\frac{1}{2}\left[\frac{1}{\alpha}F_y+F_x\right]-\frac{i}{\alpha^2}uv
\right)\psi+\frac{i}{\alpha^2}\left(\alpha u_x+u_y\right)\phi,}\\
{}\\
\displaystyle
{\phi_t=-2i\phi_{xx}+\frac{2i}{\alpha}v\psi_x+
\left(\frac{1}{2}\left[\frac{1}{\alpha}F_y-F_x\right]+\frac{i}{\alpha^2}uv
\right)\phi+\frac{i}{\alpha^2}\left(\alpha v_x-v_y\right)\psi,}
\end{array}
\eqno(44)
$$
where $q=-iF_x$.

Let $\{\psi_1,\,\phi_1;\,\psi_2,\,\phi_2;\,\psi,\,\phi\}$ are
solutions of (44) for the same	$u$, $v$ and $F$.
DT is given by
$$
\begin{array}{l}
\psi\to \psi_1=\psi_x-a\psi-b\phi,\qquad \phi\to \phi_1=\phi_x-c\psi-d\phi,\\
{}\\
u\to u_1=u+2\alpha b,\qquad
v\to v_1=v-2\alpha c,\qquad
F\to F_1=F+4i(a+d),
\end{array}
\eqno(45)
$$
where
$$
\begin{array}{l}
\displaystyle
{a=\frac{\psi_{1,x}\phi_2-\psi_{2,x}\phi_1}{\Delta},\qquad
b=\frac{\psi_{2,x}\psi_1-\psi_{1,x}\psi_2}{\Delta},\qquad
c=\frac{\phi_{1,x}\phi_2-\phi_{2,x}\phi_1}{\Delta}},\\
{}\\
\displaystyle
{d=\frac{\phi_{2,x}\psi_1-\phi_{1,x}\psi_2}{\Delta},\qquad
\Delta=\psi_1\phi_2-\psi_2\phi_1.}
\end{array}
\eqno(46)
$$
The quantities $a$, $b$, $c$ ¨ $d$ are solutions of the system
$$
\begin{array}{l}
a_y=\alpha(a_x+2 bc)+uc-vb,\qquad
b_y=\alpha(b_x+2 bd)+(d-a)u+u_x,\\
{}\\
c_y=-\alpha(c_x+2ac)+(a-d)v+v_x,\qquad
d_y=-\alpha(d_x+2bc)-uc+vb,\\
{}\\
\displaystyle
{a_t=2i\left[\left(a_x+2bc+a^2+\frac{q}{4}\right)_x+2(a+d)bc\right]+
\frac{i}{\alpha^2}\left[bv_y+cu_y-\left(uv\right)_x\right]+}\\
\displaystyle
{+\frac{i}{\alpha}\left[
cu_x-bv_x+2(cdu-abv+(uc)_x)+\frac{q_y}{2}\right],}\\
{}\\
\displaystyle
{b_t=2i\left[\left(b_x+2bd\right)_x+2\left(a_x+bc+d^2+\frac{q}{4}\right)b\right]+
\frac{i}{\alpha^2}\left[u_{xy}+(d-a)u_y-2buv\right]+}\\
\displaystyle
{+\frac{i}{\alpha}\left[u_{xx}+(3d-a)u_x+2(d_x+d^2-ad+bc)u-2b^2v\right],}\\
{}\\
\displaystyle
{c_t=-2i\left[\left(c_x+2ac\right)_x+2\left(d_x+bc+a^2+\frac{q}{4}\right)c\right]-
\frac{i}{\alpha^2}\left[v_{xy}+(a-d)v_y-2cuv\right]+}\\
\displaystyle
{+\frac{i}{\alpha}\left[v_{xx}+(3a-d)v_x+2(a_x+a^2-ad+bc)v-2c^2u\right],}\\
{}\\
\displaystyle
{d_t=-2i\left[\left(d_x+2bc+d^2+\frac{q}{4}\right)_x+2(a+d)bc\right]-
\frac{i}{\alpha^2}\left[bv_y+cu_y-\left(uv\right)_x\right]-}\\
\displaystyle
{-\frac{i}{\alpha}\left[
cu_x-bv_x+2(cdu-abv-(bv)_x)-\frac{q_y}{2}\right].}
\end{array}
\eqno(47)
$$
These equations can be obtained from the (44). Eqs. (47) are similar to the
(13)-(14). The $LA$ pair (44) and Eqs. (47) are simply two different
reprersentations of the same equations. To construct mDS equations we
must, first, express fields $a$ and $d$ via one function
$\theta=\theta(x,y,t)$ defined by the relations
$$
a=\frac{\alpha\theta_x+\theta_y}{2\alpha},\qquad
d=\frac{\alpha\theta_x-\theta_y}{2\alpha}.
\eqno(48)
$$
It is easy to check the truth of this representation by the substitution
(48) into the first and fourth equations (47). Second, we must exclude
potentials $u$ and $v$ from the first four equations (47). Introdusing
new fields  $\xi=\xi(x,y,t)$, $X=\theta_x$, $Y=\theta_y$
$$
q=-2X_x-X^2-4bc+\frac{1}{\alpha^2}\left(\xi-Y^2\right),
$$
and putting
everything together we find mDS (mDS-I if $\alpha=1$ and mDS-II if $\alpha=i$)
equations as the system
of five equations for the five functions $b$, $c$, $\xi$, $X$ and
$Y$:
$$
\begin{array}{l}
\displaystyle
{ib_t+2\left(bX+b_x\right)_x+
\frac{1}{\alpha}\left[\left(UX+U_x\right)_x+2b\left(Uc-Vb\right)-
2Y\left(bX+b_x\right)\right]+}\\
\displaystyle
{+\frac{1}{\alpha^2}\left[U_{xy}-U(X_y+XY)-2(U_xY+bUV)+\xi b\right]+
\frac{1}{\alpha^3}Y\left(UY-U_y\right)=0,}\\
{}\\
\displaystyle
{-ic_t+2\left(cX+c_x\right)_x-
\frac{1}{\alpha}\left[\left(VX+V_x\right)_x-2c\left(Uc-Vb\right)-
2Y\left(cX+c_x\right)\right]+}\\

\displaystyle
{+\frac{1}{\alpha^2}\left[V_{xy}-V(X_y+XY)-2(V_xY+cUV)+\xi c\right]-
\frac{1}{\alpha^3}Y\left(VY-V_y\right)=0,}\\
{}\\
\displaystyle
{iX_t+4\left[(bc)_x+2Xbc\right]-
\frac{2}{\alpha}\left[X(Vb-Uc)+4cb_y-2(cU)_x+bV_x-cU_x-YX_x\right]-}\\
\displaystyle
{-\frac{2}{\alpha^2}\left[(UV)_x-Vb_y-Uc_y+Y(bV+cU)\right]-
\frac{1}{\alpha^3}\left(Y^2-\xi\right)_y=0,}\\
{}\\
\displaystyle
{iY_t+4(b_yc-bc_y)+\frac{1}{\alpha}
\left[2(U_yc-Uc_y+V_yb-Vb_y)+\xi_x\right]=0,}\\
{}\\
\displaystyle
{X_y-Y_x=0,}
\end{array}
\eqno(49)
$$
where
$$
\begin{array}{l}
\displaystyle
{U=\frac{G_1}{2\alpha(2bcY+\alpha(bc_x-b_xc))}, \qquad
V=\frac{G_2}{2\alpha(2bcY+\alpha(bc_x-b_xc))},}\\
{}\\
\displaystyle
{G_1=\alpha^3[b_xX_x-bX_{xx}+2b(2bcX+b_xc-bc_x)]+\alpha^2[2(bc_y-b_yc)-
(X_x+4bc)Y]b+}\\
\displaystyle
{+\alpha(bX_{yy}-b_xY_y)+bYY_y,}\\
{}\\
\displaystyle
{G_2=\alpha^3[c_xX_x-cX_{xx}+2c(2bcX-b_xc+bc_x)]+\alpha^2[2(bc_y-b_yc)+
(X_x+4bc)Y]c+}\\
\displaystyle
{+\alpha(cX_{yy}-c_xY_y)-cYY_y}.
\end{array}
$$

As we have mentioned above, the NLS equation has two types of discrete
symmetries: Darboux and Schlesinger transformations. In just the same way,
the discrete symmetries of the DS equations include T-symmetry. This
transformations are given by ([12], [13])
$$
\begin{array}{l}
\displaystyle
{u\to u_1=u\left(uv+\alpha^2(\log u)_{xx}-(\log u)_{yy}\right),\qquad
v\to v_1=\frac{1}{u},\qquad q\to q_1=q+4(\log u)_{xx}}\\
{}\\
\displaystyle
{u\to u_{-1}=\frac{1}{v},\qquad
v\to v_{-1}=v\left(uv+\alpha^2(\log v)_{xx}-(\log v)_{yy}\right),\qquad
q\to q_{-1}=q+4(\log v)_{xx}.}
\end{array}
$$
The similar transformations for the (49) are given by

$$
\begin{array}{l}
\displaystyle
{b\to b_1=\frac{M_1}{\alpha U(U+2\alpha b)},\qquad c\to c_1=\frac{b}{U(U+2\alpha b)},}\\
{}\\
\displaystyle
{X\to X_1=X+2\alpha\frac{Ub_x-U_xb}{U(U+2\alpha b)},\qquad
Y\to Y_1=\frac{-U^2Y+2\alpha[UU_x-bU_y+\alpha U(b_x+bX)]}{U(U+2\alpha b)},}\\
{}\\
\displaystyle
{\xi\to\xi_1=\xi+4\alpha\frac{\alpha^2U(b_{xx}+(bX)_x)+\alpha
[b(UY_x-U_{xy})+U(U_{xx}-b_xY)]-UU_xY}{U(U+2\alpha b)}},
\end{array}
$$
and
$$
\begin{array}{l}
\displaystyle
{c\to c_{-1}=\frac{M_{-1}}{\alpha V(V-2\alpha c)},\qquad
b\to b_{-1}=\frac{c}{ V(V-2\alpha c)},}\\
{}\\
\displaystyle
{X\to X_{-1}=X-2\alpha\frac{Vc_x-V_xc}{ V(V-2\alpha c)},\qquad
Y\to Y_{-1}=\frac{V^2Y+2\alpha[VV_x-cV_y-\alpha V(c_x+cX)]}{ V(V-2\alpha c)},}\\
{}\\
\displaystyle
{\xi\to\xi_{-1}=\xi-4\alpha\frac{\alpha^2V(c_{xx}+(cX)_x)-\alpha
[c(VY_x-V_{xy})+V(V_{xx}-c_xY)]-VV_xY}{V(V-2\alpha c)}},
\end{array}
$$
where
$$
\begin{array}{l}
M_1=\alpha^3\left[U^2b(2X_x-X^2)+2U((U_xb-Ub_x)X+\left(U^2\right)_xb_x+
4(Ub)^2c+\left(U_x^2-2UU_{xx}\right)b\right]+\\
+\alpha^2U\left[U(UX_x+2bY_x)+(2U_yb-UU_x)X-2Ub_xY+4U^2cb-2U_{xy}b-UU_{xx}+
2b_xU_y+2U_x^2\right]+\\
+\alpha\left[U^3(Y_x+XY+Uc)+U^2(bY^2-2U_xY-U_{xy})+U_y\left((U^2)_x-bU_y\right)
\right]+U^2Y(UY-U_y),\\
{}\\
M_{-1}=\alpha^3\left[V^2c(X^2-2X_x)-2V(V_xc-Vc_x)X-\left(V^2\right)_xc_x-
4(Vc)^2b-\left(V_x^2-2VV_{xx}\right)c\right]+\\
+\alpha^2V\left[V(VX_x+2cY_x)+(2V_yc-VV_x)X-2Vc_xY+4V^2bc-2V_{xy}c-VV_{xx}+
2c_xV_y+2V_x^2\right]-\\
-\alpha\left[V^3(Y_x+XY+Vb)+V^2(cY^2-2V_xY-V_{xy})+V_y\left((V^2)_x-cV_y\right)
\right]+V^2Y(VY-V_y).
\end{array}
$$
It is possible to verify that
$X_{\pm 1,y}=Y_{\pm 1,x}$, $(b_1)_{-1}=(b_{-1})_1=b$ and so on.

To obtain dressing chains of discrete symmetries we must exclude
potentials  $u$, $v$, $q$ from the two systems: (47) and and the
system of equations obtained from (47) by replacing
 $a\to a_1$, $b\to b_1$, $c\to c_1$,
$d\to d_1$, $q\to q_1=q+4(a+d)_x$, $u\to u_1$, $v\to v_1$
(it is necessary to use (45)). As a result, we have y-chains
$$
\begin{array}{l}
\displaystyle
{[2bc(d-a)+b_xc-bc_x]\left(a_{1,y}-\alpha a_{1,x}\right)+
(bc_1-b_1c)(a_y-\alpha a_x)_x+(bc_1-b_1c)(bc_y-b_yc)+}\\
\displaystyle
{+\alpha[(b_1c+2b_1c_1+bc_1)(bc_x-b_xc)+4bc\left(ac_1(b+b_1)-b_1d(c+c_1)\right)]+}\\
\displaystyle
{+[(bc_1+b_1c)(a-d)+b_1c_x-b_xc_1](a_y-\alpha a_x)=0,}\\
{}\\
\displaystyle
{(bc_1-b_1c)[(b_1-b)_y-\alpha(b+b_1)_x]+(a-a_1-d+d_1)[\alpha(a_{1,x}b-a_xb_1)+
a_yb_1-a_{1,y}b]+}\\
\displaystyle
{+2\alpha[(b+b_1)(abc_1+cb_1d_1)-bb_1(c+c_1)(a_1+d)]=0,}
\end{array}
$$
$$
\begin{array}{l}
\displaystyle
{(b_1c-bc_1)[(c_1-c)_y+\alpha(c+c_1)_x]+(a-a_1-d+d_1)[\alpha(a_{1,x}c-a_xc_1)+
a_yc_1-a_{1,y}c]-}\\
\displaystyle
{-2\alpha[(c+c_1)(a_1c_1b+cdb_1)-cc_1(b+b_1)(a+d_1)]=0,}\\
{}\\
\displaystyle
{(a_1+d_1)_y-\alpha(a_1-d_1)_x=0}
\end{array}
\eqno(50)
$$
and  t-chains:
$$
\begin{array}{l}
\displaystyle
{\alpha^2(a_1-a)_t+i[2\alpha^2((a^2-a_1^2-2b_1c_1-a_{1,x}-d_x)_x+
2((a+d)bc-(b+b_1)c_1d_1-(c+c_1)a_1b_1-}\\
\displaystyle
{-(bc_1)_x)-b_xc_1-b_1c_x)+\alpha((c-3c_1)U_x+(b+b_1)V_x+2(b_1c_y-b_yc_1-
(a+d)_{xy}+}\\
\displaystyle
{+(b_x+a_1b_1-ab)V-(c_{1,x}+c_1d_1-cd)U))+(b-b_1)V_y+(c-c_1)U_y]=0,}\\
{}\\
\displaystyle
{\alpha^2(bb_{1,t}-b_tb_1)+i[2\alpha^2(2b^2(a_1d_1-b_1c_1)-2bb_1^2(c_1+c)+
b_{xx}b_1-b_{1,xx}b-2bd_{1,x}(b+b_1)+}\\
\displaystyle
{+2(b_xdb_1-b_{1,x}d_1b)+b[a_1b_x-2a_{1,x}b_1-b_{xx}-3b_xd_1+2(b_1(d^2-bc)-
d_1^2(b+b_1))])+}\\
\displaystyle
{+\alpha((b_1-b)U_{xx}+(3(b_1d-bd_1)+a_1b-ab_1)U_x+2[b_1bV(b+b_1)+bb_y(a_1-d_1)-
bb_{xy}+}\\
\displaystyle
{+U(b_1d_x-bd_{1,x}+b_1d^2-bd_1^2-bb_1(c+c_1)+ba_1d_1-b_1ad)])+U_{xy}(b_1-b)+}\\
\displaystyle
{+U_y(b(a_1-d_1)-b_1(a-d))]=0,}\\
{}\\
\displaystyle
{\alpha^2(cc_{1,t}-c_tc_1)+i[2\alpha^2(2c^2(b_1c_1-a_1d_1)+2cc_1^2(b_1+b)+
c_{1,xx}c-c_1c_{xx}+2ca_{1,x}(c+c_1)+}\\
\displaystyle
{+2(ca_1c_{1,x}-c_1ac_x)+c[2c_1d_{1,x}-d_1c_x+c_{xx}+3c_xa_1+2(c_1(bc-a^2)+
a_1^2(c+c_1))])+}\\
\displaystyle
{+\alpha((c_1-c)V_{xx}+(3(ac_1-a_1c)+cd_1-c_1d)V_x+2[c_1cU(c+c_1)+cc_y(d_1-a_1)-
cc_{xy}+}\\
\displaystyle
{+V(c_1a_x-ca_{1,x}+c_1a^2-ca_1^2-cc_1(b+b_1)+ca_1d_1-c_1ad)])+V_{xy}(c-c_1)+}\\
\displaystyle
{+V_y(c(a_1-d_1)-c_1(a-d))]=0,}\\
{}\\
\displaystyle
{\alpha^2(d_1-d)_t+i[2\alpha^2((d_1^2-d^2+2b_1c_1+a_x+d_{1,x})_x-
2((a+d)bc-(b+b_1)c_1d_1-(c+c_1)a_1b_1-}\\
\displaystyle
{-(b_1c)_x)+b_xc_1+b_1c_x)+\alpha((b-3b_1)V_x+(c+c_1)U_x+2(c_1b_y-b_1c_y-
(a+d)_{xy}+}\\
\displaystyle
{+(c_x+c_1d_1-cd)U-(b_{1,x}+a_1b_1-ab)V))+(c_1-c)U_y+(b_1-b)V_y]=0.}
\end{array}
\eqno(51)
$$
These formulas have awful sight! Unfortunately, I don't know how to write
them in more compact form in our gauge. (51) has the best form which
I could imagine.

Although this equations are long and difficult, the end result is quite
simple. We can see that all of
$(1+1)$ NLS formalism carried over directly into the
$(2+1)$ DS dressing chains. In particular, the chains
(51) involve $LA$ pair for the mDS (49). We do not give it here.

As noted in Sec. 3, the KP equations admit two types of chains, which
we call conjugate. It is a new result, which is characteristic precisely of
multidimensional systems. Of course, it is truth for the DS equations.

To construct the chains which are conjugate to (50)-(51), we must introduce
the new $LA$ pair for the DS equations ($p=p(x,y,t)$, $f=f(x,y,t)$):
$$
\begin{array}{l}
p_y=\alpha p_x-vf,\qquad
f_y=-\alpha f_x-up,\\
{}\\
\displaystyle
{p_t=-2ip_{xx}+\frac{2i}{\alpha}vf_x-
\left(\frac{1}{2}\left[\frac{1}{\alpha}F_y+F_x\right]-\frac{i}{\alpha^2}uv
\right)p+\frac{i}{\alpha^2}\left(\alpha v_x+v_y\right)f,}\\
{}\\
\displaystyle
{f_t=2if_{xx}+\frac{2i}{\alpha}up_x-
\left(\frac{1}{2}\left[\frac{1}{\alpha}F_y-F_x\right]+\frac{i}{\alpha^2}uv
\right)f+\frac{i}{\alpha^2}\left(\alpha u_x-u_y\right)p,}
\end{array}
\eqno(52)
$$
two matrix functions
$$
\Psi=\left(\begin{array}{cc}
\psi_1&\psi_2\\
\phi_1&\phi_2
\end{array}\right),\qquad
\Phi=\left(\begin{array}{cc}
p_1&f_1\\
p_2&f_2
\end{array}\right)
$$
and one-form
$$
d\Omega=
\Phi\Psi dx+\alpha\Phi\sigma_3\Psi dy+
2i\left(\Phi\sigma_3\Psi_x-\Phi_x\sigma_3\Psi+\frac{1}{\alpha}
\Phi\sigma_3\Psi\right),\qquad
\Omega=\int d\Omega.
$$
This one -form is closed if
$\psi_{1,2}$, $\phi_{1,2}$ , $p_{1,2}$ and
$f_{1,2}$ are solutions of Eqs. (44) and (52)  for the same
$u$, $v$ ¨ $F$. It is easy to see that the quantities
$$
\begin{array}{l}
\displaystyle
{A=\frac{p_{1,x}f_2-p_{2,x}f_1}{\Delta},\qquad
B=\frac{f_{1,x}f_2-f_{2,x}f_1}{\Delta},\qquad
C=\frac{p_{2,x}p_1-p_{1,x}p_2}{\Delta},}\\
{}\\
\displaystyle
{D=\frac{f_{2,x}p_1-f_{1,x}p_2}{\Delta},\qquad
\Delta=p_1f_2-p_2f_1.}
\end{array}
$$
are solutions of the $y$-system
$$
\begin{array}{l}
A_y=\alpha(A_x+2 BC)+uC-vB,\qquad
B_y=-\alpha(B_x+2AB)+(D-A)u-u_x\\
{}\\
C_y=\alpha(C_x+2 CD)+(A-D)v-v_x,\qquad
D_y=-A_y+\alpha(A-D)_x.
\end{array}
\eqno(53)
$$
Here, we don't need $t$-system (four equations for the
$A_t$, $B_t$, $C_t$ and $D_t$).

The Darboux transformation (45) (for the system (44)) give us the
transformation rule for the functions $A$, $B$, $C$ ¨ $D$
$$
A\to a_1=\Lambda_{11}-a,\qquad
B\to b_1=\Lambda_{12}-b,\qquad
C\to c_1=\Lambda_{21}-c,\qquad
D\to d_1=\Lambda_{22}-d,
$$
where $a$, $b$, $c$ and $d$ are defined in (46), and $\Lambda_{ik}$ are
elements of the matrix $\Lambda=\Psi\Omega^{-1}\Phi$, $i,k=1,2$.
Transforming the system (53) and exluding potentials
$u$ and $v$ from this system and from the first four equations (47) we get
$y$-chains which are conjugate to (50). There are four equations, two of
them are equaivalent to the first and fourth equations (50). The rest two
equations are given by
$$
\begin{array}{l}
\displaystyle
{(bc_1-b_1c)[(b+b_1)_y+\alpha(b+b_1)_x]+(a+a_1-d-d_1)[\alpha(a_xb_1-a_{1,x}b)+
a_{1,y}b-a_yb_1]-}\\
{}\\
\displaystyle
{-2\alpha[(b+b_1)(abc_1+a_1b_1c)-bb_1(c+c_1)(d+d_1)]=0,}\\
{}\\
{}\\
\displaystyle
{(b_1c-bc_1)[(c+c_1)_y-\alpha(c+c_1)_x]+(a+a_1-d-d_1)[\alpha(a_xc_1-a_{1,x}c)+
a_{1,y}c-a_yc_1]+}\\
{}\\
\displaystyle
{+2\alpha[(c+c_1)(bc_1d_1+b_1cd)-cc_1(b+b_1)(a+a_1)]=0.}
\end{array}
$$

The conjugate t-chains can be obtained by the same way. We do not give them
here.

\section {On the localized solutions of the DS equations}

Discrete symmetries are a good way to obtain exact solutions of
the nonlinear integrable equations. In [14] we applied the DT
(S-symmetry) to construct exact solutions to the DS-I
and DS-II equations (the same to the BLP equations see in [15]).
In particular, we have obtained the dromion solutions
of the DS-I equations. In this section we use the T-symmetry to construct
nonsingular solutions to the DS-I that fall off according to the exponential
and/or rational low along {\em all} directions in the plan. In addition
to the known dromion, we give several new solutions with above
properties. In rest path of this section we present (via S-symmetrie)
novel exact solution of
the DS-II equations describing the soliton on the plane wave background.
\newline
1. {\em DS-I equations.}

To study the DS-I equations it is convenient to use the following change
of variables:
$$
\partial_x\to\frac{1}{\sqrt{2}}\left(\partial_x+\partial_y\right),\qquad
\partial_y\to\frac{1}{\sqrt{2}}\left(\partial_x-\partial_y\right),\qquad
v\to -\frac{\bar{u}}{2}.
$$

We are interested of localized solutions to the DS-I equations which move
without shape distortions and we look for these solutions in the form
$$
u(t,x,y)=\overline {v(t,x,y)}=U(\xi,\eta)\exp(i\theta),\qquad
q(t,x,y)=Q(\xi,\eta),
$$
where
$$
\xi=x-2at,\qquad \eta=x-4bt,\qquad \theta=ax+by-(a^2+b^2)t,
$$
and $U(\xi,\eta)$ is the real function. It was shown in [16] that a solutions to the
DS-I equations can be found in this setting from the nonlinear
Liouville equation
$$
\partial_x\partial_y\Phi=\frac{1}{2}\exp(\Phi),
\eqno(54)
$$
for the function  $\Phi\equiv\log(U^2)$. This result was obtained via the
T-symmetries (see the previous section).

It is convenient to introduce the boundary conditions for the $U(x,y)$ as
$$
U(x,0)=A(x),\qquad U(0,y)=B(y),\qquad A(0)=B(0)=C,
\eqno(55)
$$
where $A(x)$ and $B(y)$ are the given functions. Now we can use the solution
to the Gousat problem for (54) given in [17], from which we find
$$
U(x,y)=\frac {4CA(x)B(y)}{4C^2-\int_0^x dp\,A^2(p)\int_0^y dq\,B^2(q)}.
\eqno(56)
$$
Using T-symmetries, one can also find the explisit expression for the
$Q(x,y)$ field
$$
\begin{array}{l}
\displaystyle {Q(x,y)=\frac{1}{2}(P'(\frac {1}{P'})''+G'(\frac {1}{G'})'')-
\frac {3}{4}([(\log P')']^2+[(\log G')']^2)+}\\
\displaystyle{+\frac {2}{PG-4}(PG''+P''G-\frac {(PG')^2+(P'G)^2+8P'G'}{PG-4})},
\end{array}
$$
where
$$
P(x)=\frac {1}{C}\int_0^x dp\,A^2(p),
\qquad G(y)=\frac {1}{C}\int_0^y dq\,B^2(q).
$$

Expression (56) also allows us to construct localized solutions. For
example, choosing the boundary functions in the form of two solutions
of the NLS equation,
$$
A(x)=\frac {a}{\cosh(\alpha x)},\qquad B(y)=\frac {a}{\cosh(\beta y)},\eqno(57)
$$
we obtain the well-known "one-dromion" solution [18]. Other localized
objects are found by choosing the boundary conditions appropriately.
Thus, assuming $A(x)$ and $B(y)$ to be solutions of the KdV equation,
we obtain a new solution that decays exponentially in {\bf all directions},
$$
\displaystyle{
U(x,y)=\frac {4a\alpha\beta}
{(\cosh(\alpha x)\cosh(\beta y))^2[4\alpha\beta-a^2\tanh(\alpha x)\tanh(\beta y)
(1-\frac {\tanh^2(\alpha x)}{3})(1-\frac {\tanh^2(\beta y)}{3})]}},
\eqno(58)
$$
where $a$, $\alpha$ and $\beta$ are real constants satisfying the condition
$a^2<9\alpha\beta$, which guarantees the nonsingular behaviour of (58). Let
us note that if if the functions $A(x)$ and $B(y)$ have maxima $m$ and
$l$, respectively, the solution $U(x,y)$ determined by Eq. (56) corresponds
to $ml$ localized bound formations. It possible to show that these
equations are different from the (L,M)-dromions [19] and $N^2$-soliton
solutions built in [20] and [21].

The rational localized solutions ("lumps") are obtained in the same way. let
us consider, for example, the following boundary conditions:
$$
\begin{array}{l}
\displaystyle{
A_1(x)=\frac {\alpha_1}{x^2+a^2},\qquad A_2(x)=\frac {\alpha_2}{x^4-(ax)^2+a^4},}\\
\displaystyle{
B_1(y)=\frac {\beta_1}{y^2+b^2},\qquad B_2(y)=\frac {\beta_2}{y^4-(by)^2+b^4}}.
\end{array}
\eqno(59)
$$
Substituting $A_1$ and $B_1$ into (56) and using (55),
one obtains the one-lump solution,

\begin{eqnarray*}
\lefteqn{{u}= - 16{C}\,{a}^{2}\,{b}^{2} \left/ {\vrule
height0.37em width0em depth0.37em} \right. \! \! ({C}^{2}\,{b}^{2
}\,{a}^{2}\,{y}\,{x} - 16\,(\,{x}^{2} + {a}^{2}\,)\,(\,{y}^{2} +
{b}^{2}\,)} \\
 & & \mbox{} + {a}\,{b}\,{C}^{2}\,(\,{S}\,{Q}\,(\,{x}^{2} + {a}^{
2}\,)\,(\,{y}^{2} + {b}^{2}\,) + {y}\,{b}\,{S}\,(\,{x}^{2} + {a}
^{2}\,) + {x}\,{a}\,{Q}\,(\,{y}^{2} + {b}^{2}\,)\,)),
\end{eqnarray*}
\[
{S}={\rm arctan} \left( \! \,{\displaystyle \frac {{x}}{{\it a}
}}\, \!  \right),\qquad
{Q}={\rm arctan} \left( \! \,{\displaystyle \frac {{y}}{{\it b}
}}\, \!  \right).
\]
Substituting $A_1$ and $B_2$ (or $A_2$ and $B_1$), we get the two-lamp
solution, and substituting $A_2$ and $B_2$, the four-lamp solution. One
can always choose the paramerers involved in (59) such that the solutions
falling off in all directions according to rational law are nonsingular.
Finally, choosen $A(x)$ from (57) and $B_1(y)$ from (59), we obtain
the localized solution falling off according to the exponential
law as a function of $x$ and  according to the ratianal
law as a function of $x$
\newline
2. {\em DS-II equations.}

We couldn't find a localized nonstationary solution of the DS-II equation.
We present only soliton on the plane wave background. We choose the initial
solution of the DS-II equations as
$$
u=A \exp\left(iS\right),\qquad S=-(2A^2+a^2-b^2)t+ax+by, \qquad q=0.
$$
where
$A$, $a$ and $b$  are real constants. The solutions of the $LA$ pair (44)
$$
\displaystyle
{\psi_1=f \exp{\left(i\frac{S}{2}+M\right)},\qquad
\psi_2=\overline{\frac{i(2m-b)-p}{2A}
\left(f-2i\frac{\alpha_1}{p}\right)}
\exp{\left(i\frac{S}{2}+\overline{M}\right)},}
$$
where
$$
\begin{array}{l}
\displaystyle
{f=\alpha_1\left[x+\frac{1}{p}\left((b-2m)y+2\left(2(bm-A^2-2m^2)
-ap\right)t\right)\right]+
\alpha_2,}\\
{}\\
\displaystyle
{M=mx+\frac{1}{2}\left((p-a)y+[p(b+2m)-4am]t\right),\qquad
p^2+\left[4A^2+(b-2m)^2\right]^2=0.}
\end{array}
$$
Using (45) (with $\overline{m}=m$, $\overline{\alpha_1}=\alpha_1$ and
$\alpha_2=0$) we get nonsingular solution $u_1$ which is the
soliton on the plane wave background:
$$
\mid u_1\mid^2\to A^2 \qquad {\rm at}\qquad x^2+y^2\to\infty
$$
This solution is  two-dimensional generalization of "exulton" solution
of the NLS built in [1].

\section{Conclision}

The past years of intense theoretical research have made it increasingly clear
that the secret to integrability most likely lies in the power of discrete
symmetries.  Let us summarize some of the promising  features of the discrete symmetries:
\newline
\newline
(1) It is the an extremely power method to construct exact solutions of integrable equations.
\newline
(2) This approach includes all	"soliton miracles": finit-gap solutions, the
Painleve property ets.
 \newline
(3) Discrete symmetries allow us to proliferate integrable equations. For
example,  the whole MKdV
theory can be consider as the DT theory and the
Toda-Volterra theory can be consider as the theory of S- and T-symmetries
of the NLS equations [9].
\newline
(4) Discrete symmetries led to connection between integrable system and
supersymmetry.
\newline
(5) Discrete symmetries allow us to construct discrete integrable systems. \newline
\newline

Ideally, we would want a unified theory to unite and understand
all "soliton miracles". There are two ways to find this unified theory.
The first path is connected with the Hirota bilinear difference
equation [22]. This famous equation is known to provide a canonical integrable
discretization for most important types of soliton equation.

The second path is  the theory of dressing chains. This approach allow us
to proliferate integrable equations and, at the same time, to establish
a links between known integrable equations (it is clear that the Miura map
can be obtained from the dressing chains). Ideally, we
want to show  that all integrable equations are
nothing but different forms of a single equation!

Even if this direction is right, it is still a long way off. Now the theory
itself often seems like a confused jumble of random (but usefull) rules and
random (but remarkable) observations. It remains to be seen how useful
the dressing chains will  become in the future.
\newline
\newline
{\bf Acknowledgement}
\newline

This work was supported by RFBR, Grant 00-01-00783 and
the Grant of Education Department of the Russian Federation,  No. E00-3.1-383.
$$
{}
$$
\centerline{\bf References}
\noindent
\begin{enumerate}
\item V.B. Matveev and M.A. Salle, {\em Darboux Transformations and Solitons},
Springer, Berlin (1991).
\item B.A. Borisov and S.A.Zykov, Theor. Math. Phys., {\bf 115}, 530 (1998).
\item A.V. Mikhailov, A.B. Shabat and R.I. Yamilov, Russ. Math. Surv.,
{\bf 42}, 1 (1987).
\item A.V. Yurov, Theor. Math. Phys., {\bf 119}, 731 (1999).
\item V.A. Andreev and M.V. Burova, Theor. Math. Phys., {\bf 85}, 376 (1990).
\item V.G. Konopelchenko and V.G. Dubrovsky, Stud. Appl. Math.,
{\bf 86}, No.3, 219 (1992).
\item A.P. Veselov and A.B. Shabat, Funct. Anal. Appl., {\bf 27},
10 (1993).
\item A.B. Shabat, Theor. Math. Phys., {\bf 121}, 165 (1999).
\item €.B. Shabat and R.I. Yamilov, Algebra and Analis., {\bf 2},
183 (1990).
\item A. Davey and K. Stewartson, Proc. R. Soc. A, {\bf 338}, 101 (1974).
\item K. Nishinari, K. Abe and J. Satsuma, Theor. Math. Phys., {\bf 99},
745 (1994).
\item A.V. Yurov, Theor. Math. Phys., {\bf 109}, 1508 (1996).
\item A.N. Leznov, A.B. Shabat, R.I. Yamilov.,
Phys. Lett. A., {\bf 174}, 397 (1993).
\item S.B.Leble, M.A. Salle and A.V. Yurov, Inverse Problems, {\bf 8}, 207
(1992).
\item A.V. Yurov, Phys. Lett. A., {\bf 262}, 445 (1999).
\item A.V. Yurov,  Theor. Math. Phys., {\bf 112}, 1113 (1997).
\item G.P. Dzhordzhadze (Jorjadze),  Theor. Math. Phys., {\bf 41}, 867 (1979).
\item  M. Boiti, J.J.-P. Leon, L. Martina and F. Pempinelli, Phys. Lett. A,
{\bf 132}, 432 (1988).
\item A.S. Fokas and P.M. Santini, Phys. Rev. Lett., {\bf 63}, 1329 (1989).
\item  M. Boiti, L. Martina, O.K. Pashaev and F. Pempinelli,
"Dinamics of Multidimensional Solitons", Lecce preprint (June 1991).
\item J. Hietarinta and R. Hirota, Phys. Lett. A, {\bf 237}, 145 (1990).
\item R. Hirota, J. Phys. Soc. Japan, {\bf 50}, 3785 (1981).
\end{enumerate}

\vfill
\eject

\end{document}